
------------------------------ Start of body part 2

%
%
%
%

\magnification=1200
for double spacing

\font\gross=cmbx10 scaled \magstep2
\font\mittel=cmbx10 scaled\magstep1
\font\pl=cmssq8 scaled \magstep1
\font\sc=cmcsc10

\font\matbf=cmmib10

\def\RR{\rm I\!R}

\def\h#1{{\cal #1}}
\def\a{\alpha}
\def\b{\beta}
\def\g{\gamma}
\def\d{\delta}

\def\l{\lambda}
\def\m{\mu}
\def\n{\nu}

\def\s{\sigma}
\def\om{\omega}
\def\na{\nabla}

\def\sq{\Square}
\def\square#1{\mathop{\mkern0.5\thinmuskip\vbox{\hrule
    \hbox{\vrule\hskip#1\vrule height#1 width 0pt\vrule}\hrule}
    \mkern0.5\thinmuskip}}
\def\Square{\mathchoice{\square{6pt}}{\square{5pt}}
    {\square{4pt}}{\square{3pt}}}


{\nopagenumbers
\null
\vskip-1.5cm
\hskip5cm{ \hrulefill }
\vskip-.55cm
\hskip5cm{ \hrulefill }
\vskip1.5mm
\hskip5cm{ {\pl  University of Greifswald, (June, 1994)}}
\vskip0.1mm
\hskip5cm{ \hrulefill }
\vskip-.55cm
\hskip5cm{ \hrulefill }
\bigskip
\hskip5cm{\ hep-th/9509077}

\vfill

\centerline{\gross The heat kernel approach}
\medskip
\centerline{\gross for calculating the effective action}
\medskip
\centerline{\gross in quantum field theory and quantum gravity}
\bigskip
\bigskip
\centerline{{\mittel I. G. Avramidi}
\footnote{*}{ Alexander von Humboldt Fellow}
\footnote{$\S$}{On leave of absence from Research Institute for Physics,
Rostov State University, Stachki 194, Rostov-on-Don 344104, Russia}
\footnote{\dag}{Talk given in Institute for Theoretical Physics, University of
Jena, Jena, Germany, at June 30, 1994}}

\centerline{\it Department of Mathematics, University of Greifswald}
\centerline{\it Jahnstr. 15a, 17489 Greifswald, Germany}
\centerline{\sl E-mail: avramidi@math-inf.uni-greifswald.d400.de}

\bigskip
\smallskip
\vfill
\centerline{\sc Abstract}
\bigskip
{\narrower
A short informal overview about recent progress in the calculation of the
effective action in quantum gravity is given. I describe briefly the standard
heat kernel approach to the calculation of the effective action and discuss the
applicability of the Schwinger - De Witt asymptotic expansion in the case of
strong background fields. I propose a new ansatz for the heat kernel that
generalizes the Schwinger - De Witt one and is always valid. Then I discuss
the general structure of the asymptotic expansion and put forward
some approximate explicitly covariant methods for calculating the heat kernel,
namely, the high-energy approximation as well as the low-energy one. In both
cases the explicit formulae for the heat kernel are given.
{\bigskip}}
\eject}


First of all I would like to thank Dr. M. Basler and Prof. Kramer for their
kind invitation and hospitality expressed to me in the University of Jena. It
is a great pleasure for me to present this talk here in Jena where a lot of
first class research in gravitation has been done.

However, I will not concern the more or less well studied classical gravity but
will try to give an overview about some recent research in {\it quantum}
gravity. More precisely, I will describe some methods of calculations that
turned out to be very effective and powerful in quantum field theory and gauge
theories.

I start with some introductory notes on quantum field theory, namely, the {\it
formal definition} of the generating function and that of the effective action.
Then I will try to explain how this formal definition becomes meaningful in
{\it perturbation theory} after {\it regularization}. The most convenient way
to do this is to employ the {\it heat kernel} approach and the $\zeta$-function
regularization. In what follows I will talk mainly about the various
approximations used in covariant calculations of the effective action, namely
the {\it Schwinger - De Witt asymptotic expansion} as well as so called {\it
high-energy} approximation and the {\it low-energy} one.

\bigskip
\bigskip
\centerline{\bf 1. Formal scheme of quantum field theory}
\bigskip
So, let us consider first some set of fields
$\varphi^i\equiv\varphi^A(x)=\{\varphi^a(x),\psi^i(x),\h B^c_\m,h_{\m\n},...\}$
defined on some, say, asymptotic flat Riemannian manifold $M$ with the metric
of the Minkovski signature $(-+\cdots+)$. The classical dynamics of these
fields is described by the classical equations of motion
$$
S_{,i}\equiv{\d S \over \d \varphi^A(x)}=0
\eqno(1)
$$
derived from the classical action functional
$$
S(\phi)=\int_M {\cal L}\left(\varphi(x),\partial_\m\varphi(x)\right)
\eqno(2)
$$.

After the quantization the fields become field operators $\hat \varphi$ acting
on some state vectors of a Hilbert space, so called Fock space. Let us suppose
that there exist some initial $|in>$ and final $<out|$ vacuum states defined in
some appropriate way. At least in asymptotically flat manifolds this can be
done consistently.

What is really important is that the vacuum-vacuum transition amplitude can be
presented as a Feynman path integral
$$
<out|in>\equiv Z(J)=\int {\cal
D}\varphi\exp\left\{i\left(S(\varphi)+J_i\varphi^i\right)\right\}
\eqno(3)
$$

Here we introduced the classical sources $J_i$ to investigate the linear
reaction of the system on the external perturbation. All the content of quantum
field theory with all quantum effects is contained in the following Green
functions
$$
\left<\hat\varphi^{i_1}\cdots\hat\varphi^{i_n}\right>\equiv
{\left<out|T\left(\hat\varphi^{i_1}\cdots\hat\varphi^{i_n}\right)\right|in>
\over <out|in>}
\eqno(4)
$$
where $T$ means the chronological ordering operator, i.e. the fields must be
arranged from left to right in order of decreasing time arguments.

It is easy to show that all these Green functions can be obtained by the
functional differentiation of the functional $Z(J)$, that is called, therefore,
the generating functional. Moreover, factorizing the unconnected contributions
one comes to generating functional for connected Green functions $W(J)=-i\log
Z(J)$
$$
\left<\hat\varphi^{i_1}\cdots\hat\varphi^{i_n}\right>
=i^{-n}\exp(-iW){\d^n\over \d J_{i_1}\cdots \d J_{i_n}}\exp(iW).
\eqno(5)
$$

The lowest connected Green functions have special names: the mean field
$$
<\varphi^i>\equiv\phi^i(J)={\d W\over \d J_i}
\eqno(6)
$$
and the propagator
$$
\left<\left(\hat\varphi^i-\phi^i\right)\left(\hat\varphi^k-\phi^k\right)\right>
\equiv -i{\cal G}^{ik}(J)=-i{\d^2 W\over \d J_i \d J_k}
\eqno(7)
$$

Further, the connected Green functions are expressed in terms of vertex
functions. The generating functional for vertex functions is defined then by
the Legendre functional transform
$$
\Gamma(\phi)=W(J(\phi))-J_k(\phi)\phi^k
\eqno(8)
$$

This is the most important object in quantum field theory. It contains all the
information about the quantized fields.
\item{1)} First, one can show that it satisfies the equation
$$
\Gamma_{,i}(\phi)\equiv {\d \Gamma\over \d \phi^i}=-J_i(\phi)
\eqno(9)
$$
and, therefore, gives the effective equations for the mean field. These
equations replace the classical equations of motion and describe the effective
dynamics of background fields taking into account all quantum corrections! Thus
$\Gamma(\Phi)$ is called usually the {\it effective action}.
\item{2)} Second, it determines the full or exact propagator of quantized
fields
$$
-\Gamma_{,ik}{\cal G}^{kn}=\d^n_i
\eqno(10)
$$
where $\d^n_i\equiv\d^B_A\d(x-y)$,
and the vertex functions
$$
\Gamma_n\equiv\Gamma_{,i_1\cdots i_n}, \qquad (n\ge 3)
\eqno(11)
$$
This means that any $S$-matrix amplitude, or any Green function, is expressed
in terms of propagator and vertex functions that are determined by the
effective action.
\item{3)} At last, when the test sources vanish the effective action is just
the vacuum amplitude
$$
<out|in>\Big\vert_{J=0}=\exp\left(i\Gamma\right)\big\vert_{J=0}
\eqno(12)
$$
It determines then the probability that the $out$-vacuum is still a vacuum and
does not contain $in$-particles
$$
|<out|in>|^2\Big\vert_{J=0}=\exp\left(-2{\rm Im} \Gamma\right)\big\vert_{J=0}
\eqno(13)
$$
Therefore, the imaginary part of the effective action determines the total
number of created particles.

\bigskip
\bigskip
\centerline{\bf 2. Perturbation theory}
\bigskip

Let us rewrite the definition of the effective action. To give more sense to
the path integral it is convenient to make a so called Wick rotation, or
Euclidization, i.e. one replaces the real time coordinate to the purely
imaginary one $x^0 \to ix^0$ and singles out the imaginary factor also from the
action $S \to iS$ and the effective action $\Gamma \to i\Gamma$. Then the
metric of the Riemannian manifold becomes Euclidean, i.e. positive definite,
and the classical action in all {\it good} field theories becomes positive
definite functional.
So, the Euclidean effective action is defined to satisfy the equation
$$
\exp\left(-{1\over \hbar}\Gamma(\phi)\right)=
\int {\cal D}\varphi\exp\left\{-{1\over \hbar}
\left(S(\varphi)-(\varphi^i-\phi^i)\Gamma_{,i}(\phi)\right)\right\}
\eqno(14)
$$

This path integral is still not well defined. There is not any reasonable
method, except for lattice theories, to calculate this integral in general
case. The only path integrals that can be well defined are the Gaussian ones
$$
\int {\cal D}h\exp\left\{-{1\over 2}
\left(h^iF_{ik}h^k\right)\right\}
={\cal N}({\rm Det}F)^{-1/2}
\eqno(15)
$$
$$
{\int {\cal D}h\exp\left\{-{1\over 2}
\left(h^iF_{ik}h^k\right)\right\}h^{j_1}\cdots h^{j_n}\over
\int {\cal D}h\exp\left\{-{1\over 2}
\left(h^iF_{ik}h^k\right)\right\}}
=\cases{{(2m)!\over m!}G^{(j_1j_2}\cdots G^{(j_{2m-1}j_{2m})},  &\qquad n=2m\cr
0 &\qquad n=2m+1}
\eqno(16)
$$

Here ${\rm Det}F$ is the functional determinant of the operator $F$, ${\cal N}$
is some inessential, actually infinite, constant that could be put, in
principle equal to 1, and
$G^{ik}=(F_{ik})^{-1}$ is the Green function of the operator $F$ with Euclidean
boundary conditions, i.e it must be regular and bounded at the infinity. Thus
the full path integral can be well defined as an asymptotic series of Gaussian
ones. This is just the quasiclassical, or WKB, approximation in the usual
quantum mechanics. We decompose the fields in the classical and quantum parts
$$
\varphi=\phi+\sqrt\hbar h
\eqno(17)
$$
and look for a solution of the equation for the effective action in form of an
asymptotic series in powers of Planck constant.
$$
\Gamma(\phi)=S(\phi)+\sum\limits_{n \ge 1}\hbar^n \Gamma_{(n)}(\phi)
\eqno(18)
$$
Then all the coefficients of this expansion can be found. They are expressed in
terms of the well-known Feynman diagrams. The number of loops in these diagrams
correspond to the power of the Planck constant. We will be interested below in
the so called {\it one-loop effective action}
$$
\Gamma_{(1)}={1\over 2}\log{\rm Det} F
\eqno(19)
$$
where $F_{ik}=S_{,ik}$.

\bigskip
\bigskip
\centerline{\bf 3. $\zeta$-function regularization}
\bigskip

Although this quantity seems to be very easy it is still ill defined. The point
is it is divergent. This is just the well-known ultraviolet divergence of the
quantum field theory. Indeed, we can rewrite the functional determinant as
$$
\Gamma_{(1)}={1\over 2}\log \prod_n\limits \l_n ={1\over 2}\sum\limits_n\log
\l_n
\eqno(20)
$$
where $\lambda_n$ are the eigenvalues of the operator $F$. This series is easy
to show to be divergent.

That means that one needs a regularization. This point was investigated very
thoroughly by many authors and it is found that in quantum gravity and gauge
theories the most appropriate regularizations are the analytical ones. The
functional determinants can be well defined in terms of the so called
$\zeta$-function.
It is defined by
$$
\zeta (p)=\mu^{2p}{\rm Tr} F^{-p}={\mu^{2p}\over \Gamma(p)}\int\limits_0^\infty
dt t^{p-1}{\rm Tr} U(t)
\eqno(21)
$$
Here $U(t)$ is the so called {\it heat kernel} and Tr$U(t)$ is the functional
trace of it
$$
{\rm Tr} U(t)=\int dx {\rm tr}U(t)\Big\vert_{diag}
\eqno(22)
$$
$$
U(t)\Big\vert_{diag}=\exp(-tF)\d(x,x')\Big\vert_{x=x'}
\eqno(23)
$$

It is easy to show that the integral over $t$ converge for sufficiently large
$Re p$. In the rest of the complex plane of $p$ the $\zeta$-function should be
defined by analytic continuation.
This analytic continuation leads then to a meromorphic function with some poles
on real axis. But the most important point is that it is analytic at the point
$p=0$. This means that one can calculate the values of the $\zeta$-function and
its derivatives at the point $p=0$. Formally $\zeta(0)$ is equal to the total
number of modes of the operator $F$
$$
\zeta(0)={\rm Tr} 1=\sum\limits_n 1
\eqno(24)
$$
i.e. one can say that $\zeta(0)$ counts the modes of the operator $F$.

It is almost obvious that the formal derivative of the $\zeta$-function is
equal to
$$
\zeta'(0)=-{\rm Tr}\log{F\over \m^2}=-\log{\rm Det}{ F\over \m^2}
\eqno(25)
$$
where $\zeta'(p)=d\zeta(p)/dp$, i.e. the functional determinant of the operator
$F$. Here $\mu$ is a dimensionful parameter which is introduced to preserve
dimensions. It is called usually renormparameter.

So, the one-loop effective action becomes now a well defined object
$$
\Gamma_{(1)}=-{1\over 2}\zeta'(0)
\eqno(26)
$$
Moreover, studying the change of the renormparameter
$$
\m{\partial\over \m}\zeta(p)=2p\zeta(p)
\eqno(27)
$$
$$
\m{\partial\over \m}\Gamma_{(1)}=-\zeta(0)
\eqno(28)
$$
one can get the explicit dependence of the effective action on $\mu$
$$
\Gamma_{(1)}(\m)=-\zeta(0)\log{\m\over M} + \Gamma_{(1)}(M)
\eqno(29)
$$
where $M$ is some fixed mass or energy parameter which cab be chosen from some
physical grounds.

\bigskip
\bigskip
\centerline{\bf 4. Heat kernel for operators of Laplace type}
\bigskip

Everything said so far was rather formal. Now we will be more concrete. As you
have seen the most important thing in one-loop calculations is the heat kernel.
For a very wide class of field theories it is sufficient to consider the second
order differential operator of Laplace type, namely
$$
F=-\sq + Q + m^2
\eqno(30)
$$
where
$$
\sq=g^{\mu\nu}\nabla_\mu\nabla_\nu
=g^{-1/2}(\partial_\m+{\cal A}_\m)g^{1/2}g^{\m\n}(\partial_\n+{\cal A}_\n)
\eqno(31)
$$
is the Laplacian (or Dalambertian in hyperbolic case), $Q=\{ Q^A_{\
B}(x)\}$ is a matrix-valued potential term,
${\cal A}_\mu=\{{\cal A}^A_{\ B\m}(x)\}$ is an arbitrary connection and
$m$ is a mass parameter. Every second order operator with leading symbol given
by metric tensor can be put in this way.

So, the heat kernel is defined now by requiring it to satisfy the heat equation
$$
\left({\partial\over\partial t}+F\right)U(t\vert x,x^\prime)=0
\eqno(32)
$$
with the following initial condition
$$
U(0\vert x,x^\prime)=g^{-1/2}(x)\delta (x,x^ \prime)
\eqno(33)
$$

To solve this equation we single out first the asymptotic factor
$$
U(t\vert x,x')=(4\pi t)^{-d/2}\Delta^{1/2}(x,x')
\exp\left(-{\sigma(x,x')\over 2t}\right){\cal P}(x,x')\Omega (t\vert x,x')
\eqno(34)
$$
Here $\sigma(x,x')$ is the geodetic interval defined as one half the square of
the length of the geodesic between points $x$ and $x'$, $\Delta(x,x')$ is the
Van Vleck - Morette determinant and ${\cal P}(x,x')$ is the parallel
displacement operator. The trace of the heat kernel that defines
$\zeta$-function and the effective action is expressed in terms of the transfer
function $\Omega(t)$
$$
{\rm Tr}U(t)=(4\pi t)^{-d/2}{\rm Tr}\Omega(t)
\eqno(35)
$$
which satisfies the equation
$$
\left({\partial\over\partial t }
+{1\over t}D+\bar F\right)
\Omega(t\vert x,x')=0
\eqno(36)
$$
with initial condition
$$
\Omega(0\vert x,x^\prime )\bigg\vert_{x=x^\prime}=1
\eqno(37)
$$
where $D=\na_\mu\sigma^\mu$ is the differential operator along the geodesic and
$\bar F$ is defined by
$$
\bar F={\cal P}^{-1}\Delta^{-1/2}F\Delta ^{1/2}{\cal P}
=-{\cal P}^{-1}\Delta^{-1/2}\sq\Delta ^{1/2}{\cal P}
+{\cal P}^{-1}Q{\cal P} + m^2
\eqno(38)
$$

The usual way to solve this equation is to employ so called Schwinger - De Witt
ansatz {\sc De Witt} (1965)
$$
\Omega(t)=\exp(-tm^2)\sum_{k=0}^{\infty}{(-t)^k\over {k!}}a_k
=\sum_{k=0}^{\infty}{(-t)^k\over {k!}}b_k
\eqno(39)
$$
where
$$
b_n=\sum\limits_{k=0}^n {n\choose k} m^{2(n-k)}a_k
\eqno(40)
$$
Then one gets the well known De Witt recursion relation for the coefficients
$a_k$
$$
\left(1+{1\over k}D\right)a_k=\bar F_0a_{k-1}
\eqno(41)
$$
where
$$
\bar F_0=\bar F\big\vert_{m=0}
=-{\cal P}^{-1}\Delta^{-1/2}\sq\Delta ^{1/2}{\cal P}
+{\cal P}^{-1}Q{\cal P}
\eqno(42)
$$
These coefficients play very important role both in physics and mathematics. We
call them Hadamard - Minackshisundaram - De Witt - Seeley (HMDS) coefficients.
The calculation of these coefficients in general case of arbitrary background
is in itself of great importance and offers a complicated technical problem. Up
to now only four lowest order coefficients, more precisely their coincidence
limits were calculated. The pioneering method of {\sc De Witt} (1965) is quite
simple but gets very cumbersome at higher orders. By means of it only two
coefficients $a_1, a_2$ were calculated. The approach of mathematicians differs
considerably from that of physicists. It is very general but also very
complicated and seems not to be well adopted to physical problems. It allowed
to compute in addition the third coefficient $a_3$ {\sc Gilkey} (1975).

It is the general manifestly covariant technique for the calculation of the
coefficients $a_k$ that was elaborated in my PhD thesis in Moscow University in
1987. This technique is essentially based on the use of the covariant Taylor
expansions of all needed two-point quantities, such as the second derivatives
of the geodesic interval and the first derivative of the parallel displacement
operator. I solved the De Witt recursion relations explicitly and obtained a
very effective nonrecursive covariant formulae for coefficients $a_k$. This
method allowed me to calculate the coefficient $a_4$ as well as to maintain the
previous results of other authors. These results are published in {\sc
Avramidi} (1989, 1990, 1991).  The fourth coefficient $a_4$ for the case of
scalar operators was also calculated by {\sc P. Amsterdamski, A. L. Berkin and
D. J. O'Connor} (1989).

Moreover, it allows to analyze the general structure of $a_k$ and to calculate
all of them in some approximation, for example, the terms with leading
derivatives in all $a_k$.

One should mention here that the Schwinger - De Witt ansatz is actually an
asymptotic expansion and not a Taylor series. This expansion does not converge,
in general, namely in the case when the transfer function $\Omega(t)$ is not
analytic at the point $t=0$.

\bigskip
\bigskip
\centerline{\bf 5. New ansatz for the heat kernel}
\bigskip

Therefore, I proposed in {\sc Avramidi} (1990, 1991) a {\it new} ansatz for the
transfer function $\Omega(t)$ which generalizes the Schwinger - De Witt one but
is always valid. One can show that the transfer function can be always
presented in the form of an inverse Mellin transform of a product of the
$\Gamma$-function and an {\it entire} function $b_q$
$$
\Omega(t)={1\over 2\pi i }\int\limits_{c-i\infty}^{c+i\infty}dqt^q
\Gamma(-q)b_q
\eqno(43)
$$
with $c<0$.
The function $b_q$ must satisfy in addition some asymptotic conditions at $q\to
\pm i\infty$ so that this integral converge at the infinity and one can simply
move the contour of integration.

Then if we move the integration contour to the right, so that it cross the real
axis between $N-1$ and $N$ than the residues at simple poles of the
$\Gamma$-function reproduce exactly the first $N$ terms of the Schwinger - De
Witt expansion
$$
\Omega(t)=\sum_{k=0}^{N-1}{(-t)^k\over {k!}}b_k
+{1\over 2\pi i} \int\limits_{c_{_N}-i\infty}^{c_{_N}+i\infty}dq
t^q \Gamma(-q)b_q
\eqno(44)
$$
where $N-1<c_{_N}<N$.

This formula is exact, in contrast to the original Schwinger - De Witt
expansion. This means that it gives the exact form of the rest term of the
Schwinger - De Witt expansion which can be evaluated from some independent
grounds.
Substituting this new ansatz in the equation for the transfer function we
obtain the recursion relation for the function $b_q$
$$
\left(1+{1\over q}D\right)b_q=\bar F b_{q-1}
\eqno(45)
$$
with initial condition
$$
b_0=1
\eqno(46)
$$

So, the function $b_q$:
\item{1)} contains the whole information about the heat kernel and, therefore,
about the effective action,
\item{2)} is the analytic continuation of the HMDS-coefficients on the whole
complex plane of their order, so that it is just $b_k$ at integer positive
points $q=k$,
\item{3)} it provides the evaluation of the rest term of the Schwinger - De
Witt asymptotic expansion.

Thus the recursion equation for $b_q$ is actually equivalent to the original
heat equation.
Using the function $b_q$ we can get also a {\it new} ansatz for the complex
power of the differential operator $F$
$$
\eqalignno{
F^{-p}=&{1\over \Gamma (p)}\int\limits_0^\infty dt\ t^{p-1}U(t) &\cr
=&(4\pi)^{-d/2}{\Delta^{1/2}\over 2\pi i\Gamma(p)}
\int\limits_{c-i\infty}^{c+i\infty}dq
\left({\sigma\over 2}\right)^{p-{d\over 2}+q}
\Gamma(-q)\Gamma\left(-p+{d\over 2}-q\right)b_q &   (47)\cr}
$$
with $c<{d\over 2}-{\rm Re} p$.
This form is especially useful in analyzing the short-distance behavior of the
Green function $F^{-1}$ when $x \to x'$ or $\s \to 0$.

Now taking the coincidence limit $x\to x'$ one can get the functional trace of
the complex power of the differential operator $F$, i.e. the $\zeta$-function
$$
\zeta (p)=(4\pi)^{-d/2}{\Gamma\left(p-{d\over 2}\right)\over \Gamma(p)}
\mu^{2p}B_{{d\over 2}-p}
\eqno(48)
$$
This is a very simple but very important formula. It displays the whole
analytical structure of the $\zeta$-function. Because $B_q$ is an entire
function it is immediately seen that $\zeta(p)$ is  meromorphic function and
all poles of it are given simply by the $\Gamma$-function. Anyway $\zeta(p)$ is
analytic at $p=0$ and one can calculate simply the values of $\zeta$-function
and its derivative at this point.

One has to distinguish here between the spaces of even and odd dimension. In
even dimension  $\zeta(0)$ is just the HMDS-coefficient of order $d/2$
$$
\zeta(0)=(4\pi)^{-d/2}{(-1)^{d/2}\over \Gamma(d/2+1)}B_{d/2},
\eqno(49)
$$
where
$$
B_q={Tr}b_q= \int dx\ g^{1/2}{\rm tr}b_q\big\vert_{diag}
\eqno(50)
$$
while in odd dimension it vanishes
$$
\zeta(0)=0.
\eqno(51)
$$
The one-loop effective action is given by the formulas:
for even dimension
$$
\Gamma_{(1)}={1\over 2}(4\pi)^{-d/2}
{(-1)^{d/2}\over \Gamma\left({d\over 2}+1\right)}
\left\{B'_{d/2}-\left[\log\mu^2+\Psi\left({d\over 2}+1\right)+{\matbf C}\right]
B_{d/2}\right\}
\eqno(52)
$$
where $\Psi(q)=(d/dq)\log\Gamma(q), \ {\matbf C}=-\Psi(1)$ and
$$
B'_{d/2}={d\over dq}B_q\bigg\vert_{q=d/2}
\eqno(53)
$$
and for odd dimension
$$
\Gamma _{(1)}={1\over 2}(4\pi)^{-d/2}
{\pi(-1)^{d-1\over 2}\over \Gamma\left({d\over 2}+1\right)}
B_{d/2}
\eqno(54)
$$
The nontrivial contribution to the one-loop effective action is contained here
in the first terms which do not depend on the renormalization ambiguity, i.e.
on the renormparameter $\mu$.
These terms do not depend also on the regularization. The remaining term
$B_{d/2}\log \mu$ (in even dimension) is just the renormalization one. We are
free to add such terms to the effective action because they are different in
different regularization schemes. The models that do not have such terms, i.e.
when $\Gamma_{(1)}$ does not depend on the renormparameter $\mu$ are called
one-loop finite. The examples of such models are: odd dimensional and
sypersymmetric models. This does not mean, of course, that these models are
completely finite. To state this one has to prove independence on the
renormparameter at all orders of the perturbation theory.

\bigskip
\bigskip
\centerline{\bf 6. General structure of the asymptotic expansion}
\bigskip

Now after I spent so much time to convince you how important it is to calculate
the one-loop effective action, $\zeta$-function and the heat kernel I am going
to describe very briefly some methods for the calculation of these quantities
and just present the results not going deeply into details.

Let me make some remarks on the subject.
\item{$\bullet$}
First of all, it is obviously impossible to evaluate the effective action
exactly, even at the one-loop order. There are, of course, some simple special
cases of background fields and geometries that allow the exact and even
explicit calculation of the heat kernel or the effective action. However, the
effective action is an {\it action}, i.e. a functional of background fields
that should be varied to get the Green functions, the vacuum expectation values
of various fields observables, such as energy-momentum tensor and Yang-Mills
currents etc.. That is why one needs the effective action for {\it general}
background and, therefore, one has to develop consistent {\it approximate}
methods for its calculation.
\item{$\bullet$}
Second, in quantum gravity and gauge theories the effective action is a {\it
covariant} functional, i.e. invariant under diffeomorphisms and gauge
transformations. That is why the approximations for calculating the effective
action have to preserve the {\it general covariance at each order}. The flat
space perturbation theory is an example of bad approximation because it is not
covariant.

\bigskip\bigskip
\leftline{ \sl 6.1 Massive quantum fields in weak background fields}
\bigskip

One of the most known and succesful covariant approximations is that of {\it
weak background fields}, i.e. in the case when the Compton wave length of the
massive field is much smaller than the characteristic length scale $L$ of the
background fields
$$
{\hbar\over mc} \ll L
\eqno(55)
$$
In other words, all invariants build from the curvature and its covariant
derivatives are much smaller than the corresponding power of the mass parameter
$$
R\ll m^2, \qquad \na\na R\ll m^4, \dots
\eqno(56)
$$
that means
$$
a_k \ll m^{2k}
\eqno(57)
$$

In this case one can simply use the Schwinger - De Witt ansatz for the heat
kernel to get the $1/m^2$ - asymptotic expansion of the effective action: for
odd $d$
$$
\Gamma _{(1)}={1\over 2}(4\pi )^{-d/2}\pi (-1)^{d-1\over 2}
\sum\limits_{k=0}^{\infty}{m^{d-2k}\over k!\Gamma
\left({d\over 2}-k+1\right)}A_k
\eqno(58)
$$
and for even dimension
$$
\Gamma_{(1)}={1\over 2}(4\pi )^{-d/2}\Biggl\{(-1)^{d/2}
\sum\limits _{k=0}^{d/2}
{m^{d-2k}\over k!\Gamma \left({d\over 2}+1-k\right)}
A_k\left[\ln {m^2\over \mu^2}-
\Psi \left({d\over 2}-k+1\right) -{\matbf C} \right]
$$
$$
+\sum\limits_{k={d\over 2}+1}^{\infty}
{\Gamma \left(k-{d\over 2}\right)(-1)^k\over k!m^{2k-d}}A_k\Biggr\}
\eqno(59)
$$

This is a good approximation in weak background fields and describes the
physical effect of vacuum polarization of the massive fields. For example, in
four-dimensional space this expansion looks like
$$
\Gamma_{(1)}={1\over 2}(4\pi )^{-2}\Biggl\{
{1\over 2}m^4\left(\log{m^2\over \m^2}-{3\over 2}\right)A_0+
m^2\left(\log{m^2\over \m^2}-1\right)A_1
+\log{m^2\over \m^2}A_2
$$
$$
-{1\over 6 m^2}A_3+{1\over 24 m^4}A_4+O({1\over m^6})\Biggr\}
\eqno(60)
$$
where $A_0, A_1$ and $A_2$ are the renormalization terms and $A_3\sim R^3$ and
$A_4\sim R^4$ describe really the vacuum polarization effects. Variation of the
$\Gamma_{(1)}$ with respect to the background metric gives then the vacuum
expectation value of the energy-momentum tensor, variation with respect to the
connection gives the vacuum ecpectation value of the Yang-Mills current etc.

However the Schwinger - De Witt approximation is of very limited applicability.
It is not effective in the case of large or rapidly varying background fields
and becomes meaningless in massless theories. Therefore, this approximation can
not describe essentially nonperturbative nonlocal effects such as particle
creation and the vacuum polarization by strong background fields.

\bigskip\bigskip
\leftline{ \sl 6.2 General structure of HMDS-coefficients and partial
summation}
\bigskip

In fact, the effective action is a nonlocal and nonanalytical functional and
posseses a sensible massless limit. But its calculation requires quite
different methods. One such method which would exceed the limits of the
Schwinger - De Witt asymptotic expansion is the {\it partial summation
procedure} {\sc Vilkovisky} (1984). It is based on the analysis of the general
structure of the HMDS-coefficients. The HMDS-coefficients are the local
polynomial invariants
built from the curvature and its covariant derivatives. Therefore, one can
classify all the terms according to the number of curvatures and their
derivatives. The terms with leading derivatives can be shown to have the
followin structure $R\sq^{k-2}R$.
Then it follows the class of terms with one more curvature etc. The last class
of terms does not contain any covariant derivatives at all but only the powers
of the curvature
$$
\eqalignno{
A_k &= \int dx g^{1/2}{\rm tr}\,\Biggl\{
        R \sq^{k-2} R +
       \sum_{ 0\le i \le 2k-6} R\ \na^i R\ \na^{2k-6-i}R &\cr
       &+ \cdots + (\na R)R^{k-3}(\na R)+R^k\Biggr\} &(61)\cr}
$$

Now one can try to sum up each class of terms separately to get the
corresponding expansion of the heat kernel
$$
\eqalignno{
{\rm Tr}U(t)=&\int dx g^{1/2}(4\pi t)^{-d/2}\exp(-tm^2){\rm tr}
\Biggl\{1-t\left(Q-{1\over 6}R\right)		&\cr
&+t^2R\chi(t\sq)R+\cdots+
t^3\na R\Psi(tR)\na R + \Phi(tR)\Biggr\}	&(62)\cr}
$$
and that of the effective action
$$
\Gamma_{(1)}=\int dx g^{1/2}{\rm tr}
\Biggl\{R F(\sq)R+\cdots+\na R Z(R)\na R + V(R)\Biggr\}
\eqno(63)
$$

We mention here once again that these expansions are asymptotic ones and do not
converge, in general. So, one has to use some methods of summation of the
divergent asymptotic series. This can be done by using an integral transform
and analytic continuation.

Let us stress again that in quantum gravity and gauge theories the assumptions
about the local behavior of the background fields must deal with physical gauge
invariant properties of the local geometry, i.e. the curvature invariants but
not with the behavior of the metric and the connection which is not invariant.
Comparing the value of the curvature with that of its {\it covariant}
derivatives one comes to two possible approximations.

\bigskip
\bigskip
\centerline{\bf 7. High-energy approximation}
\bigskip

\def\nohkf {
    + {{t^2}\over {2}}\biggl[Q\gamma^{(1)}(t\Square)Q+
     2\h R_{\alpha\mu}\nabla^\alpha{{1}\over {\Square}}\gamma^{(2)}
     (t\Square)\nabla_\nu\h R^{\nu\mu}
          -2Q\gamma^{(3)}(t\Square)R  }
\def\nohks {+R_{\mu\nu}\gamma^{(4)}
      (t\Square)R^{\mu\nu}
              +R\gamma^{(5)}(t\Square)R\biggr]  }
\def\bodd {
             \int\limits_0^1d\xi\,f^{(i)}(\xi)
             \left(m^2-{{1-\xi^2}\over {4}}\Square\right)
             ^{{{d}\over {2}}-2 } }

There are here two general approximations. The first one is the so called
high-energy approximation or the short-wave one. It is characterized by small
but rapidly varying curvature. It means that the covariant derivatives of the
curvatures are much greate than the powers of them
$$
\na\na R \gg RR
\eqno(64)
$$
Such a formulation is manifestly covariant and, therefore, more suitable for
calculations in quantum gravity and gauge theories than the usual flat space
perturbation theory.
The terms with higher derivatives in HMDS-coefficients have the following form
$$
\eqalignno{
A_k=\int dx\,g^{1/2}{\rm tr}{(-1)^{k-2}\Gamma(k+1)\Gamma(k-1)\over
2\Gamma(2k-2)}\biggl\{
&f^{(1)}_k Q\sq^{k-2}Q
+2f^{(2)}_k \h R_{\alpha\mu}\nabla^\alpha\sq^{k-3}\nabla_\nu\h R^{\nu\mu}
-2f^{(3)}_k Q\sq^{k-2}R &\cr
&+f^{(4)}_k R_{\mu\nu}\sq^{k-2}R^{\mu\nu}+f^{(5)}_k R\sq^{k-2}R
+O(R^3)\biggr\}			&  (65)\cr}
$$
where the coefficients $f^{(i)}_k$ are given by
$$
\eqalignno{
&f^{(1)}_k=1\qquad,\qquad f^{(2)}_k={{1}\over {2(2k-1)}}\qquad,\qquad
f^{(3)}_k={{k-1}\over {2(2k-1)}}\qquad,&\cr
& f^{(4)}_k={{1}\over {2(4k^2-1)}}\qquad,\qquad
f^{(5)}_k={{k^2-k-1}\over {4(4k^2-1)}}		& (66)\cr}
$$

They were calculated also in my PhD thesis in 1987 and published then in (1989,
1990). At the same time just the same result was obtained in {\sc T. Branson,
P. B. Gilkey and B. \O rsted}(1990) by using completely independent approach.

Now having {\it all} $A_k$ and supposing that these are the main terms in the
high-energy approximation we can try to sum up the local Schwinger - De Witt
expansion to get the nonlocal heat kernel with appropriate formfactors
$$
     \eqalignno{
  {\rm Tr}\,U(t) &=
     \int dx\, g^{1/2}(4\pi t)^{-d/2}\exp{(-tm^2)}{\rm tr}\biggl\{
                1-t\left(Q-{{1}\over {6}}R\right)               & \cr
                &\nohkf                                    &\cr
                &\nohks             +O(R^3)\biggr\}\qquad . &(67)\cr}
$$
where the formfactors $\g^{(i)}(t\sq)$ have the form
$$
\gamma^{(i)}(t\Square)=\int\limits_0^1 d\xi\,f^{(i)}(\xi)
\exp\left({{1-\xi^2}\over {4}}t\Square\right)
\eqno(68)
$$
with some simple polynomial functions $f^{(i)}(\xi)$ given by
$$
\eqalignno{
&f^{(1)}(\xi)=1\qquad,\qquad f^{(2)}(\xi)={{1}\over {2}}\xi^2
\qquad,\qquad f^{(3)}(\xi)={{1}\over {4}}(1-\xi^2)\qquad, &\cr
&f^{(4)}(\xi)={{1}\over {6}}\xi^4\qquad,\qquad
f^{(5)}(\xi)={{1}\over {48}}(3-6\xi^2-\xi^4) &(69)\cr}
$$
Using this heat kernel one obtains then the $\zeta$-function and the effective
action.

Another way to get the effective action that we have put forward is much more
elegant and simple. One has first to calculate the coefficients $B_k$ which are
also {\it local} polynomials. They are just the HMDS-coefficients for
non-vanishing mass. And then one has to make a very nontrivial trick, namely,
the {\it analytic continuation} of these coefficients. In this way we obtain
the function $B_q$ without solving the recursion relations. Of course, we have
used them to calculate the HMDS-coefficients $A_k$.

Having the analytic function $B_q$ one can calculate then very simply the
$\zeta$-function and the effective action.

We stress here again that the heat kernel as well as the function $B_q$ (for
noninteger $q$) and $\zeta$-function and effective action are {\it nonlocal}
functionals of the background fields. The result for the effective action looks
like {\sc Avramidi } (1989, 1990)
$$
\Gamma_{(1)}=\Gamma_{(1)loc}+\Gamma_{(1)nonloc}
\eqno(70)
$$
Here the local part is equal: in odd $d$
$$
\Gamma_{(1)loc} ={{1}\over {2}}(4\pi)^{-d/2}{{\pi (-1)^{{{d-1}
                       \over {2}}
                        }}\over {\Gamma({{d}\over {2}} +1)}}
                        \int dx\,g^{1/2}str\left\{m^d
                        +{{d}\over {2}}m^{d-2}\left(Q-{{1}\over {6}}R\right)+
                     O(R^3)\right\}
\eqno(71)
$$
and in even dimension
$d$
$$
     \eqalignno{
     \Gamma_{(1)loc} &={{1}\over {2}}(4\pi)^{-d/2}{{(-1)^{d/2}}\over {\Gamma
                        ({{d}\over {2}}+1) }}\int dx\,g^{1/2}str\biggl\{
                        m^d\left[\ln {{m^2}\over {\mu^2}}
                        -\Psi\left({{d}\over {2}}\right)
                        -{\bf C}\right]                               & \cr
                       &+{{d}\over {2}}m^{d-2}\left[\ln{{m^2}\over {\mu^2}}
                       -\Psi\left({{d}\over {2}}\right)-{\bf C}\right]\left(Q
                       -{{1}\over {6}}R\right)
                       +O(R^3)\biggr\}     \qquad .           &(72)\cr}
$$
The nonlocal part of the effective action can be written down in the form
$$
       \eqalignno{
       \Gamma_{(1)nonloc}&={1\over 2}(4\pi )^{-d/2}\int dx \,g^{1/2}\,str
                           \biggl\{Q\beta^{(1)}(\Square)Q
                           +2\h R_{\alpha\mu}\nabla^\alpha{{1}\over {\Square}}
                           \beta^{(2)}(\Square)\nabla_\nu\h R^{\nu\mu}& \cr
                          &-2Q\beta^{(3)}(\Square)R
                          +R_{\mu\nu}\beta^{(4)}(\Square)R^{\mu\nu}
                          +R\beta^{(5)}(\Square)R
                          +O(R^3)\biggr\}                   &(73)\cr}
$$
where $\b^{(i)}(\sq)$ are {\it nonlocal} formfactors. They have the following
integral representation: for odd $d$
$$
        \beta^{(i)}(\sq)={{\pi(-1)^{(d-1)/2}}
        \over {2\Gamma({{d}\over {2}}-1)}}
        \bodd                                         \eqno(74)
$$
and for even $d$
$$
        \eqalignno{
            \b^{(i)}(\sq)=
             {{(-1)^{d/2}}\over {2\Gamma({{d}\over {2}}-1) }}
            & \bodd                                          & \cr
            & \times \left\{\ln\left[{{1}
            \over {\mu^2}}\left(m^2-
             {{1-\xi^2}\over {4}}\Square\right)\right]
             -\Psi\left({{d}\over {2}}-1\right)-{\matbf C}\right\} \qquad .
                                                             &(75)\cr}
$$
with the same functions $f^{(i)}$.

Here it is immediately seen the difference between the even and odd dimension.
The formfactors in odd dimension do not have the logarithmic term depending on
the renormparameter. This is a direct consequence of the ultraviolet finiteness
of the effective action in this case.

One can show that this integrals define analytic functions on the whole complex
plane with the cut along the positive real ais from $m^2/4$ to $\infty$. The
asymptotic of the formfactors at the infinity have the form
$$
\b(\l)\bigg\vert_{\l\to\infty}=(-\l)^{d/2-2}\left(C_1\log{-\l\over\m^2}
+C_2+O({m^2\over\l})\right)
\eqno(76)
$$
Moreover, using this integral representation one can analyse their analytic
properties, calculate their high-energy limits and imaginary parts in the
pseudo-Euclidean region above the threshold etc..

Let us consider a simple example illustrating our technique, namely the
massless conform invariant scalar field on a two-dimensional manifold. In this
case the HMDS-coefficients have a very simple explicit form
$$
B_k={k(k-1)\over 2}{(\Gamma(k+1))^2\over \Gamma(2k+2)}
\int dx g^{1/2} R\sq^{k-2}R
\eqno(77)
$$
This formula gives immediately the function $B_q$ simply by putting $k$ to be
complex. Calculating then the derivative of the function $B_q$ at the point
$q=1$ we obtain the corresponding effective action.
$$
\Gamma_{(1)}=-{1\over 2(4\pi)} B'_1
={{1}\over {24(4\pi)}}\int dx\,g^{1/2}
\left\{R{{1}\over{\Square}}R+O(R^3)\right\}
\eqno(78)
$$

This is exactly the famous {\sc Polyakov} (1981) effective action which was
obtained by using completely different method, namely, by integrating the
conformal anomaly. The point is the following. Any two-dimensional manifold is
conformally flat. Therefore, any functional of the metric can be determined by
the functional derivative with respect to conformal factor, i.e. by the so
called trace anomaly.

One has to mention here the work of {\sc Barvinsky and Vilkovisky} (1987, 1990)
and {\sc Barvinsky, Gusev, Zhitnikov and  Vilkovisky} (1993) who developed a
different method for calculating the nonlocal effective action. They managed to
calculate the next {\it third order} in the nonlocal curvature expansion.

\bigskip
\bigskip
\centerline{\bf 8. Low-energy approximation}
\bigskip

Let us go over now to the opposite {\it low-energy} or long-wave approximation,
which is used for calculating the effective potential in quantum field theory.
The low-energy effective action is determined by strong slowly varying
background fields. Therefore, it can be obtained by summing up the terms
without derivatives in the first place. It leads to a local but nonanalytical
functional. Such a result  can not be obtained by any perturbation theory and
is essentially nonperturbative.

This means that the derivatives of the curvature are much smaller than the
powers of it.
$$
\na\na R \ll R R
\eqno(79)
$$
To obtain the effective potential it is sufficent to consider the lowest
(zeroth) order of the low-energy approximation. It means that we can put all
covariant derivatives of the curvature and the potential term equal to zero
$$
\na_\m R_{\a\b\g\d} = 0,\qquad \na_\m{\cal R}_{\a\b}=0,\qquad \na_\m Q = 0.
\eqno(80)
$$
Mention that these conditions are local. The determine the geometry of the
symmetric spaces. They can be of very different topological structure. However,
we do not investigate the influence of the topology and concentrate our
attention on the local effects.

In physical problems the correct setting of the problem is as follows. Consider
a complete noncompact asymptotically flat manifold without boundary that is
homeomorphic to $\RR^d$. Let a finite region of the manifold exist  that is
strongly curved and quasi-homogenous, i.e. the local invariants of the
curvature vary very slowly. Then the geometry of this region is locally very
similar to that of a symmetric space. This can be imagined as follows. Take the
flat Euclidean space $\RR^d$, cut out from it a region $M$ with some boundary
$\partial M$ and stick to it along the boundary, instead of the piece cut out,
a piece of a curved symmetric space with the same boundary $\partial M$.

This case is more comlicated than the high-energy one. Nevertheless, the
problem turns out to be purely algebraic one.
I will briefly present here only the main idea and the results obtained in {\sc
Avramidi} (1993, 1994).

\bigskip\bigskip
\leftline{\sl 8.1 Heat kernel in flat space with nonvanishing Yang-Mills
background}
\bigskip

So, let us consider first the case of Yang-Mills background fields and flat
space, i.e. the Riemann curvature vanishes
$$
R_{\a\b\g\d}=0
\eqno(81)
$$
The operators of covariant derivatives toghether with the curvature build a
nilpotent Lie algebra
$$
\eqalignno{
[\na_\m,\na_\n]  &= \h R_{\m\n}               &(82)\cr
[\na_\m,\h R_{\a\b}] &= [\h R_{\m\n}, \h R_{\a\b}]=[\h R_{\m\n}, Q]= 0
            &(83)\cr}
$$
One can prove an algebraic theorem that expresses the heat kernel operator,
i.e. the exponential of a second order operator, in terms of the group
elements, i.e. the exponential of the first order operators.
$$
\eqalignno{
\exp(t\sq) =& (4\pi t)^{-d/2}
\det\left({tg^{-1}\h R\over \sinh(tg^{-1}\h R)}\right)^{1/2}&\cr
&\int dk
\exp\left\{-{1\over 4t}k^\m(t\h R \coth(g^{-1}t\h R))_{\m\n}k^\n +
k^\m\na_\m\right\}                              &(84)\cr}
$$

Using this representation one can act on the $\d$-function and obtain the trace
of the heat kernel
$$
{\rm Tr}U(t) = \int dx g^{1/2}(4\pi t)^{-d/2}
{\rm tr}\left\{\exp\left(-t(m^2 + Q)\right)
\det\left({tg^{-1}\h R \over \sinh(tg^{-1}\h R)}\right)^{1/2}\right\}
\eqno(85)
$$
and, therefore, the $\zeta$-function and the corresponding effective action.
This is the generalization of the well-known Schwinger's result in quantum
electrodynamics, i.e. in the case of Abelian $U(1)$ gauge group.
Here the determinant is taken over the spacetime indices and the trace over the
group ones.

This is a good example how one can get the heat kernel without solving any
differential equations.

\bigskip\bigskip
\leftline{\sl 8.2 Heat kernel for a scalar field in symmetric space}
\bigskip

The second example is more complicated and interesting. This is just a scalar
field in symmetric space, that means that the curvature ${\cal R}_{\m\n}$ is
equal to zero
$$
{\cal R}_{\m\n}=0
\eqno(86)
$$

Our main idea remains the same - to express the heat kernel operator in terms
of some exponentials of first order operators.

Let us express first the Riemann curvature of the symmetric space in the form
$$
R^{a\ c}_{\ b\ d}=\b^{ik}D^a_{\ ib}D^c_{\ kd}
\eqno(87)
$$
Here the matrices $D_i=\{D^a_{\ ib}\}$ are the generators of the isotropy
algebra
$$
[D_i, D_k] = F^j_{\ ik} D_j
\eqno(88)
$$
and $F^j_{\ ik}$ are the structure constants of it.
Matrix $\b^{ik}$ is some symmetric nondegenerate matrix. It can be treated as
the metric in the isotropy group and can be used to raise and lower the indices
of the isotropy algebra.

Then one can introduce the constants $C^A_{\ BC}=-C^A_{\ CB}$
$$
C^i_{\ ab}=E^i_{\ ab}, \quad C^a_{\ ib}=D^a_{\ ib}, \quad C^i_{\ kl}=F^i_{\
kl}, \eqno(89)
$$
$$
C^a_{\ bc}=C^i_{\ ka}=C^a_{\ ik}=0,
$$
and show that they satisfy the Jacobi identities
$$
C^E_{\ D[A}C^D_{\ BC]}=0
\eqno(90)
$$

This means that these constants are the structure constants of some Lie
algebra. This is exactly the Lie algebra of the infinitesimal isometries
$$
[\xi_A,\xi_B]=C^C_{\ AB}\xi_C
\eqno(91)
$$

Further, introducing the Cartan metric on this algebra
$$
\g_{AB} = \left(\matrix{ g_{ab} & 0             \cr
	       0                & \b_{ik}        \cr}\right). \eqno(92)
$$
one can present the Laplacian in terms of the generators of the isometries
$$
\sq = g^{\m\n}\na_\m\na_\n = \g^{AB}\xi_A\xi_B
\eqno(93)
$$
Using this representation and the structure of the algebra of isometries one
can prove a theorem, expressing the heat kernel operator in terms of an average
over the group of isometries
$$
\eqalignno{
\exp(t\sq) = (4\pi t)^{-D/2} \int d k \g^{1/2}
	&\det\left({\sinh(k^AC_A/2)\over k^AC_A/2}\right)^{1/2} &\cr
	& \times\exp\left\{ -{1\over 4t}k^A\g_{AB}k^B
	+ {1\over 6} R_G t\right\}\exp(k^A\xi_A) ,        &(94)\cr}
$$
Here
$$
R_G= -{1\over 4}\g^{AB} C^C_{\ AD}C^D_{\ BC}
\eqno(95)
$$
is the scalar curvature of the group manifold and the matrices
$C_A=\{C^B_{\ AC}\}$ and $F_i=\{F^j_{\ ik}\}$ are the generators of adjoint
representations of the algebra of isometries and the isotropy subalgebra.

Now acting with this heat kernel on the $\d$-function, one can obtain finally
the trace of the heat kernel
$$
\eqalignno{
{\rm Tr}U(t)= & \int dx g^{1/2} (4\pi t)^{-d/2}
\exp\left\{-t\left(m^2+Q-{1\over 6} R_G \right)\right\} &\cr
&\times\Bigg<\det\left({\sinh(\sqrt t \om^iF_i/2)\over \sqrt t
\om^iF_i/2}\right)^{1/2}
\det\left({\sinh(\sqrt t \om^iD_i/2)\over \sqrt t
\om^iD_i/2}\right)^{-1/2}\Bigg> 	& (96)\cr}
$$
in form of an Gaussian average over the isotropy subgroup
$$
<f(\om)> = (4\pi)^{-p/2}\int d \om \b^{1/2}
\exp\left(-{1\over 4}\om^i\b_{ik}\om^k \right) f(\om)
\eqno(97)
$$

This expression is manifestly covariant, because all its ingredients, matrices
$F_i$, $D_i$ and $\b_{ik}$ are the invariants of the curvature tensor. If we
expand this heat kernel in asymptotic series in powers of $t$ then we recover
{\it all} HMDS-coefficients for {\it all} symmetric spaces. They will be
expressed then in terms of various foldings of the quantities $F_i$, $D_i$ and
$\b_{ik}$. But these are just the curvature invariants. That means that one can
obtain the explicit formulae for HMDS-coefficients in terms of the curvature.

\bigskip
\bigskip
\centerline{\bf References}
\bigskip
\item{} {\sc P. Amsterdamski, A. L. Berkin and D. J. O'Connor}, (1989),
		Class. Quantum  Grav. {\bf 6},  1981

\item{} {\sc I. G. Avramidi}, (1987), {\it The covariant methods for
calculation of the effective action in quantum field theory and the
investigation of higher derivative quantum gravity}, PhD Thesis (Moscow State
           University, Moscow)
\item{} {\sc I. G. Avramidi}, (1989), Teor. Mat. Fiz. {\bf 79},  219
\item{} {\sc I. G. Avramidi}, (1990), Phys. Lett. {\bf B238}, 92
\item{} {\sc I. G. Avramidi}, (1989), Yad. Fiz. {\bf 49}, 1185
\item{} {\sc I. G. Avramidi}, (1990), Phys. Lett. {\bf B236}, 443
\item{} {\sc I. G. Avramidi}, (1991), Nucl. Phys. {\bf B355}, 712
\item{} {\sc I. G. Avramidi}, (1993), Yad. Fiz. {\bf 56}, 245
\item{} {\sc I. G. Avramidi}, (1993), Phys. Lett. {\bf B 305},   27
\item{} {\sc I. G. Avramidi}, (1994), {\it Covariant algebraic calculation of
the one-loop effective potential in non-Abelian gauge theory and a new approach
to	stability problem}, gr-qc/9403035, publ. in J. Math. Phys. {\bf 36} (1995)
1557
\item{} {\sc I. G. Avramidi}, (1994), {\it Covariant methods for calculating
the low-energy effective action in quantum field theory and quantum gravity},
gr-qc/9403036
\item{} {\sc  I. G. Avramidi}, (1994), {\it A new algebraic approach for
calculating the heat kernel in quantum gravity}, hep-th/9406047, J. Math. Phys.
to appear
\item{} {\sc I. G. Avramidi}, (1994), Phys. Lett. B, {\bf 336}, 171

\item{} {\sc A. O. Barvinsky, Yu. V. Gusev, V. V. Zhytnikov and G. A.
		Vilkovisky}, (1993) Covariant perturbation theory (IY), Report
		of the University of Manitoba (University of Manitoba,
		Winnipeg)
\item{} {\sc  A. O. Barvinsky and G. A. Vilkovisky}, (1985),
		Phys. Rep. {\bf C 119}, 1
\item{} {\sc  A. O. Barvinsky and G. A. Vilkovisky}, (1987), Nucl. Phys.
		{\bf B282}, 163
\item{} {\sc  A. O. Barvinsky and G. A. Vilkovisky}, (1990), Nucl. Phys.
		{\bf B333}, 471

\item{} {\sc T. Branson, P. B. Gilkey and B. \O rsted}, (1990),
		Proc. Amer. Math. Soc. {\bf 109},  437

\item{} {\sc B. S. De Witt}, (1965){\it Dynamical Theory of Groups and Fields}
           (Gordon \& Breach, New York)
\item{} {\sc B. S. De Witt}, (1975), Phys. Rep. {\bf C19}, 296
\item{} {\sc B. S. De Witt},  (1979), in {\it General Relativity}, edited by S.
Hawking and W. Israel (Cambridge University Press, Cambridge)
\item{} {\sc B. S. De Witt}, (1984), in {\it Relativity, Groups and Topology
II}, edited by B. S.  De Witt and R. Stora (North Holland, Amsterdam) p. 393

\item{} {\sc S. A. Fulling}, (1982), SIAM J. Math. Anal. {\bf 13},  891
\item{} {\sc S. A. Fulling and G. Kennedy}, (1988), Transact. Amer. Math. Soc.
{\bf 310}, 583
\item{} {\sc S. A. Fulling}, (1990), J. Symbolic Comput. {\bf 9}, 73

\item{} {\sc P. B. Gilkey}, (1975), J. Diff. Geom. {\bf 10},  601
\item{} {\sc P. B. Gilkey}, (1984), {\it Invariance Theory, the Heat Equation
and the Atiyah - Singer Index Theorem}
	    (Publish or Perish, Wilmington, Delaware (USA))

\item{} {\sc V. P. Gusynin}, (1989), Phys. Lett. {\bf B255}, 233
\item{} {\sc V. P. Gusynin}, (1990), Nucl. Phys. {\bf B333}, 296
\item{} {\sc V. P. Gusynin and V. A. Kushnir}, (1991), Class. Quant. Grav. {\bf
8}, 279

\item{} {\sc F. H.  Molzahn, T.  A.  Osborn and S. A. Fulling}, (1990), Ann.
Phys. (USA) {\bf 204}, 64

\item{} {\sc A. M. Polyakov}, (1981), Phys. Lett. B103,  207

\item{} {\sc R. Schimming}, (1981), Beitr. Anal. {\bf 15}, 77
\item{} {\sc R. Schimming}, (1990), Math. Nachr. {\bf 148}, 145
\item{} {\sc R. Schimming}, (1994), {\it Calculation of the Heat Kernel
Coefficients,} in B. Riemann Memorial Volume, edited by T. M. Rassias,
	    (World Scientific, Singapore, to be published)

\item{} {\sc J. S. Schwinger}, (1951), Phys. Rev.{\bf 82}, 664

\item{} {\sc G. A. Vilkovisky}, (1984), in {\it Quantum Theory of Gravity},
edited by S. Christensen (Hilger, Bristol) p. 169
\item{} {\sc G. A. Vilkovisky}, (1992), Class. Quant. Grav. {\bf 9}, 895
\item{} {\sc G. A. Vilkovisky}, (1992), \ {\it Heat Kernel: \ Recontre Entre
Physiciens et Mathematiciens},  Preprint CERN - TH. 6392/92,
	    in: Proc. of Strasbourg
		Meeting between physicists and  mathematicians (Publication de
		l' Institut de Recherche Math\'ematique  Avanc\'ee,
		Universit\'e  Louis
		Pasteur, R.C.P. 25, vol.43 (Strasbourg, 1992)), p. 203

\bye